\documentclass[a4paper,10pt]{article}

\usepackage{a4wide}
\usepackage{graphicx,setspace}
\usepackage{bm}
\usepackage{amsmath,textcomp}
\usepackage{amssymb}
\usepackage{subfig}
\usepackage{xcolor}

\begin{document}
\vspace*{-3cm}
\begin{flushleft}
{\Large
\textbf{Measuring the effect of node aggregation on community detection}
}
\\
Y\'{e}rali Gandica$^{1,2,\ast}$,
Adeline Decuyper$^{1}$,
Christophe Cloquet$^{1,2,3}$, 
Isabelle Thomas$^{1}$ and \\ 
Jean-Charles Delvenne$^{1,2}$
\\
\singlespacing
\footnotesize {
{1}  Center for Operations Research and Econometrics (CORE), Universit{\'e} catholique de Louvain, Louvain-la-Neuve, Belgium.\\
{2} Institute of Information and Communication Technologies, Electronics and Applied Mathematics (ICTEAM), Universit{\'e} catholique de Louvain, Louvain-la-Neuve, Belgium \\
{3} Poppy, Rue Van Bortonne, 7, 1090 Jette, Belgium.
}
\end{flushleft}

\singlespacing
\begin{abstract} 
Many times the nodes of a complex network, whether deliberately or not, are aggregated for technical, ethical, legal limitations or privacy reasons. A common example is the geographic position: one may uncover communities in a network of places, or of individuals identified with their typical geographical position, and then aggregate these places into larger entities, such as municipalities,  thus obtaining another network. The communities found in the networks obtained at various levels of aggregation may exhibit various degrees of similarity, from full alignment to perfect independence. This is akin to the problem of ecological and atomic fallacies in statistics, or to the Modified Areal Unit Problem in geography.

We identify the class of community detection algorithms most suitable to cope with node aggregation, and develop an index for aggregability, capturing to which extent the aggregation preserves the community structure. We illustrate its relevance on real-world examples (mobile phone and Twitter reply-to networks). Our main message is that any node-partitioning analysis performed on aggregated networks should be interpreted with caution, as the outcome may be strongly influenced 
by the level of the aggregation.
\end{abstract}

\singlespacing
\section*{Background}

One of the most efficient way to analyze a complex network is by partitioning the nodes into blocks, shedding light on the internal structure of the network. For example, community detection seeks to decompose a network into blocks of nodes with many internal edges, and few edges falling between the blocks. This approach allows analyzing large networks as a sum of dense but weakly interconnected subnetworks. In the last couple of decades, the wider and wider availability of various large network-shaped data has pushed the need for new formalisations to the task of community detection, and more efficient algorithms~\cite{fortunato2010}. 

Often, one same situation can be modelled by several networks of interests, where nodes represent entities at different levels of abstractions. For example, at one level a node may represent an individual person, and at another level it may represent the aggregation of several persons sharing an attribute, for example  belonging to the same age class or living in the same municipality. In this case, the edge between two aggregation classes is typically weighted as the sum of the weight of all edges linking the individuals across the two classes.  The reasons for considering an aggregated network rather than a disaggregated one are many. For instance only the aggregated network may be available to the researcher due to privacy reasons, or due to limited resources (e.g. only aggregate flows may be accessible to the measurement, or the disaggregated network may be too large to handle for a given community detection algorithm). The aggregated network may also be more relevant for a given analysis, because the aggregation removes possible noise present at the individual level, and creates statistically robust entities. 

In all these situations, it is natural to wonder whether the communities computed on different levels of aggregation will be comparable in any way. This is the question that we explore in this article. 

That some statistical patterns, for instance correlations, may differ starkly when computed either on a dataset or on an aggregated version of the same dataset is well known in statistical sciences. Extrapolating observations on categories of individuals to the individuals themselves is generically called an \emph{ecological fallacy}, with Simpson's paradox \cite{simpson1951interpretation,simpson_first_explanation} or Robinson's paradox~\cite{robinson_original_work} as well-known examples. In geographical sciences, a particular form  of such fallacy is called the  Modifiable Areal Unit Problem (MAUP). In the earliest detected occurrence of MAUP,  Gelhke and Biel \cite{gehlke1934certain} showed that the value of the correlation coefficient of geolocalised features was influenced by the size of the spatial units used in their analyses. Openshaw further showed that the results of quantitative spatial models and statistics may depend highly on the size and shape of the basic spatial units used \cite{openshaw1984modifiable}. This problem has been broadly studied and is the object of extensive literature, see \cite{wong} for a review. The \emph{atomic fallacy} can be seen as the bias generated by extrapolating patterns present at the individual level to the level of the group to which those individuals or their geographical entities belong. 

To the best of our knowledge, however, the impact of atomic and ecological fallacies on community detection has not been considered in the literature, despite its high relevance in practical applications. This is a gap that we aim to fill in the present paper, by measuring quantitatively the  impact of node aggregation on the community structure in networks. We first bring a theoretical argument showing that some community detection methods are more robust than others to node aggregation, in that whenever the communities found optimal by the method on the finer network happen to be unambiguously aggregated, the aggregated communities are also found optimal by the community detection method.
 Then, we introduce the aggregability index, a quantitative proxy for the robustness of the community structure of a given network with respect to given node aggregation classes.

We illustrate our considerations on two real-life examples. Both compare networks of places, where the nodes are geographical areas, and the edges represent interactions between areas. In these examples it is easy to generate a series of aggregated networks by merging the places into larger and larger areas, either according to an administrative hierarchy (districts, municipalities, counties, etc.) or according to coarser and coarser square grids. In the first real-life example, each node is the mean position of a Twitter user in Belgium, and edges count the reply-to tweets between two such users. Aggregated versions of this network are produced by merging the positions into larger and larger administrative units or grid cells, and merging the edges accordingly. In the second example, the nodes are mobile phone towers in and around Brussels (the capital city of Belgium), and the edges count the number of phones calls between two towers. Aggregated versions of this network are also produced by successive merging of the nodes and edges similarly to the Twitter case. The quantitative tools we introduce allow to observe that the Twitter networks exhibit significantly different community structures at different levels of aggregation, while the mobile phone networks' communities are relatively insensitive to aggregation. 

\section*{Edge-counting objective functions for optimal partitioning}

Partitioning the nodes of a network is often performed through optimising an objective function, and assigning a real number to each partition. We characterise a class of objective functions that preserve optimality of the community partition under aggregation whenever possible, as we now define.

Assume we want to detect communities in a weighted, undirected graph $G$, understood as a (non-overlapping) partition $\mathcal{C}$ of the nodes of $G$. Let us assume that we are also interested in optimising a certain criterion, capturing structural patterns of interest, typically high density of edges inside the communities and low density across communities. Some other criteria are also possible as, for instance one may want to detect core-periphery structure or general stochastic block models 
\cite{cucuringu2014detection, newman2015generalized, goldenberg2010survey}. We want to underline here that there is a variety of possible criteria whose relevance is strongly dependent on the network and the application. For instance, some methods integrate a resolution parameter that imposes a preference for small or large communities \cite{reichardt2006,delvenne2013stability}. Some methods based on comparison with a generative model for the graph are highly dependent on the choice of  the  model \cite{peel2016ground}. Even more broadly, different goals for community detection may lead to entirely different objective functions~\cite{schaub2016many,chakraborty2017metrics}. As many of those methods proceed by optimising a `goodness' criterion, we talk of \textquotedblleft the optimal partition\textquotedblright to denote the communities found to be optimal for the criterion of interest --- we suppose for simplicity that the  partition is unique and can be discovered effectively, although in practice most algorithms are only heuristics.  

Assume moreover that a graph $G'$ is obtained from the aggregation of the nodes and edges of $G$, following a partition $\mathcal{P}$ of the nodes of $G$. In other words, if $\mathcal{P}$ partitions nodes of $G$ into $k$ ``aggregation classes'',  then $G'$ has $k$ nodes. The weight of the edge (if any) between node $i$ and node $j$ of $G'$ is the sum of the weights of all edges of $G$, between nodes in the corresponding aggregation class $I$ and aggregation class $J$ of the partition $\mathcal{P}$. In particular, node $i$ of $G'$ has a self-loop aggregating the weight of all the edges inside the corresponding aggregation class $I$, representing the interactions between different nodes of the same class. In summary, the weights in $G'$ are given by
\begin{equation}
w_{ij}=\sum_{u \in I, v \in J, u \neq v} w_{uv},
\end{equation} 
where $w_{uv}$ is the weight of the edge between nodes $u$ and $v$, in the initial disaggregated network $G$. In all cases, we insist that node aggregation as considered in this paper leads to weighted aggregated graphs, even when the original graphs are unweighted.

In general, we want to understand the relationship between the communities of $G$ and $G'$. Those communities takes place on different graphs, thus a direct comparison is not possible. We can nevertheless ``lift'' the communities of $G'$ back to $G$, by replacing each node in $C'$ with its aggregation class in $G$. Indeed a community of $G'$ is a set of nodes of $G'$, each of which represents an aggregation class of $G$. To state it more formally, if $f_{\mathcal{P}}:\mathrm{Nodes}(G) \to \mathrm{Nodes}(G')$ is the aggregation function relating every node of the original graph $G$ to its corresponding node in the aggregated graph $G'$, then a community $C' \subset \mathrm{Nodes}(G')$ is lifted back to $f^{-1}(C')$, which is a notation for the set $\{x \in \mathrm{Nodes}(G) : f(x) \in C'\}$. Doing so for each community of $G'$, we obtain a partition of the nodes of $G$, which we denote $f^{-1}(\mathcal{C'})$, and which we call the ``lifting'' of $\mathcal{C'}$. This partition can now be compared to $\mathcal{C}$.

 A specific case of interest is when we want to know the communities of $G$ while we only have access to $G'$. Clearly the best scenario is when the aggregation classes are subsets of the optimal communities in $G$, i.e. when each community is a union of aggregation classes. In this case, the aggregation transforms unambiguously the optimal community partition $\mathcal{C}$ of $G$ into a (possibly non-optimal) community partition $\mathcal{C}'$ of $G'$.  From the knowledge  of the community partition $\mathcal{C}'$ of $G'$, it is then possible to recover $\mathcal{C}$ as $f^{-1}(\mathcal{C'})$, the lifting of $\mathcal{C'}$. 
  If, moreover the community structure $\mathcal{C}'$ is also optimal in $G'$ then we have a natural way to recover the community structure of the original $G$: first compute $\mathcal{C'}$ as the optimal community structure of $G'$ then lift it to $\mathcal{C}$. However, whether $\mathcal{C'}$ is indeed optimal for $G'$ depends on the criterion used to define `(optimal) communities'.

This can be guaranteed if the objective function, evaluated on a given graph $G$ and a proposed community partition $\mathcal{C}$, only depends on the graph $G''$, defined as the graph obtained by aggregating $G$ with respect to the partition $\mathcal{C}$ of the nodes. In other words, we require that the objective function depends only on the total weight of all edges between any pair of communities (including from a community to itself), but not on the way those edges are distributed inside a community or between communities. We call such a function an \emph{edge-counting function}. 

This natural result is proved simply. Since we assume that $G'$ is obtained from $G$ by aggregation with respect to a partition $\mathcal{P}$, and that the partition $\mathcal{C}$ is coarser than  $\mathcal{P}$, then the aggregation of $G'$ with respect to $\mathcal{C}'$ coincides with the aggregation of $G$ with respect to $\mathcal{C}$. Therefore, the 
edge-counting objective function takes the same value for $(G,\mathcal{C})$ and $(G',\mathcal{C}')$. Thus if $\mathcal{C}$ is optimal for $G$ then $\mathcal{C}'$ is also optimal for $G'$.

Despite its simplicity, this first result suggests that some methods of the literature are more appropriate than others in presence of node aggregation. Such edge-counting 
criteria include modularity~\cite{newman2004}, the Hamiltonian given by Potts models~\cite{reichardt2007partitioning},
linearised partition stability~\cite{delvenne2013stability}, Infomap's  description length~\cite{rosvall2008}, conductance\cite{Kannan07}, 
Normalised Cuts~\cite{Shi2000}, and their natural extension to weighted graphs. 

On the other hand, methods based on counting paths rather than edges depend  on the way edges are distributed inside a community and not only the 
number of edges or total weights. Such methods include Markov clustering\cite{vanDongen00}, Walktrap~\cite{LatapyPons08}, partition stability~\cite{delvenne2013stability}, etc., and should be used with the greatest caution in case of aggregated data. 

\section*{Different aggregations lead to different community structures}

Even an edge-counting objective function cannot preserve the community structure in the context of arbitrary aggregation classes. Assume for instance, that the aggregation classes are chosen randomly, every node being attributed uniformly randomly to one of the classes. Then, it is reasonable to assume that the aggregated graph will behave like a complete graph with all edges of similar weight. Such a graph is expected to  exhibit either no community structure, or  communities created only by the small random fluctuations in the weights, retaining no information from the optimal communities of $G$.

\begin{figure}[h!]
\begin{center}
 \includegraphics[width=0.6 \textwidth]{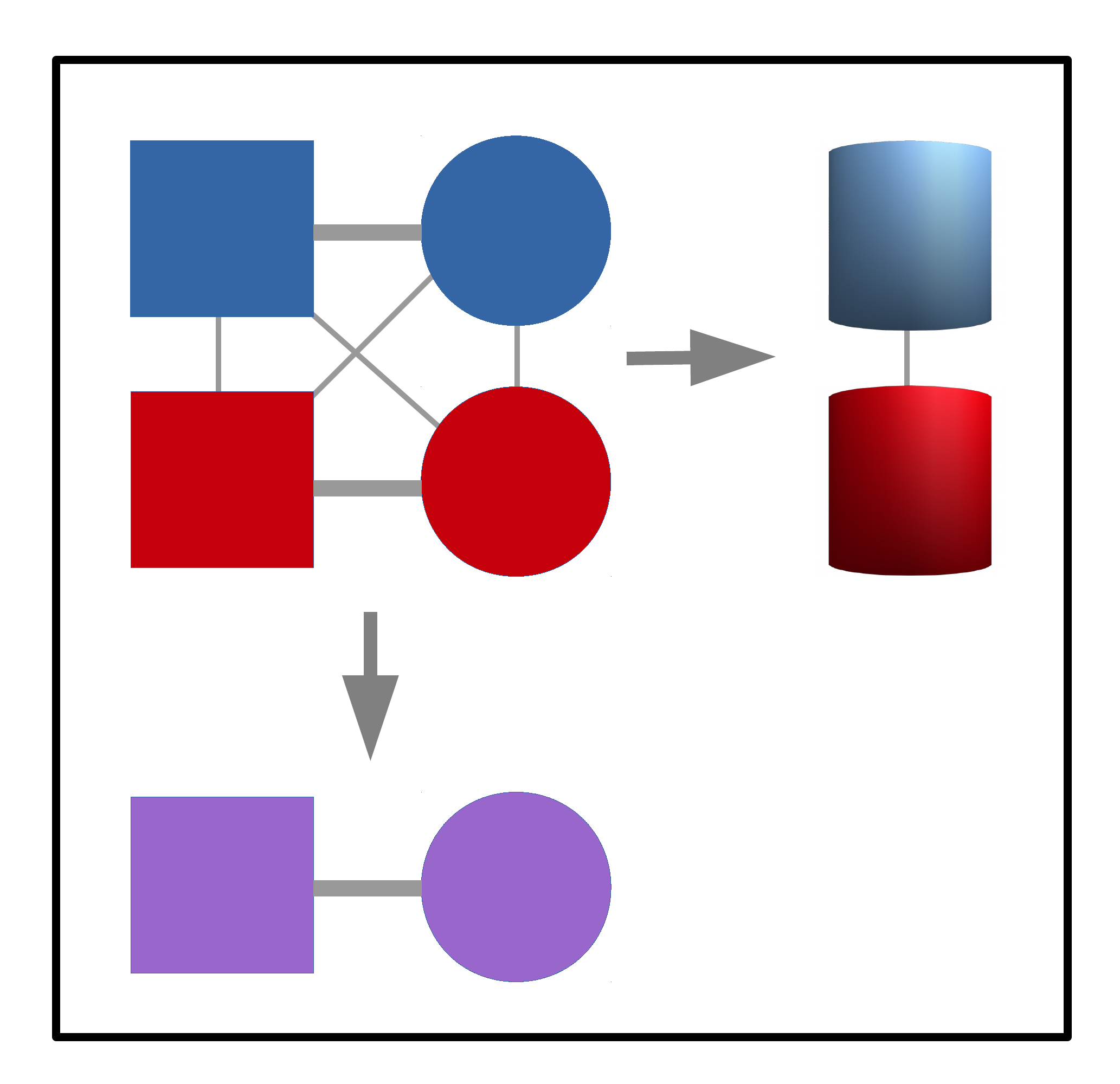}
\end{center}
\caption{Community detection over two examples of aggregations of a same 4-node network. Self-loops in aggregated networks are omitted for clarity. We assume that the community
detection criterion is such that each aggregated network admits the trivial two-community partition as optimal. The community structure on each aggregated network lifts to two 
possible partitions on the 4-node network. The community structure could be coincide with either of the two, or with that 4-community partition, according to the respective weight 
of the edges. On the depicted example, it may coincide with the same-colour communities.}  
\label{fig:toy}
\end{figure}

One can also generate examples where well chosen classes generate a graph with entirely different, yet relevant, community structure. See Fig.~\ref{fig:toy} for an illustration on a toy 4-node network, and two aggregated  2-node networks, whose communities lift back to  different community structures on the fine-scale network. These partitions may or may not coincide with the community structure computed directly on the fine-scale 4-node network---depending on the criterion for detecting communities. Here we do not specify an explicit community detection criterion, but it is reasonable to assume that if the self-loop (omitted for clarity in Fig.~\ref{fig:toy}) on each node of a 2-node  aggregated network is heavy enough compared to the internode link, then the criterion will find the two 1-node-community partition as optimal.
Suppose for example that on each aggregating partition (either by colour or by shape) of the figure, the community detection criterion is such that the 2-community partition is optimal.  Suppose as well that the community partition criterion finds the same-colour communities to be optimal on the 4-node network. We see that this partition is `orthogonal' to the partition that would be lifted from the communities on the (bottom) same-shape aggregated network. Yet all community partitions are `correct' and relevant for their respective networks: one should refrain from thinking that the aggregation leads to the `wrong' communities. 

A more general example is built with the Kronecker product of an $n_1$-node graph $G_1$ and an $n_2$-node graph $G_2$. In the product graph $G_1 \otimes G_2$, whose node set is the Cartesian product of the two individual node sets, a node $(i,j)$ is connected to the node $(i',j')$ if  $i$ and $i'$ are neighbours in $G_1$, as well as $j$ and $j'$ in $G_2$. If the graphs are weighted, then the weight on an edge in the product graph is simply the product of the weights in the corresponding edges in  $G_1$ and $G_2$.  The product graph can be aggregated in two natural ways, in one that retrieves $G_1$ as aggregated graph, and another one that retrieves $G_2$.  Assume that the fine-grained network is $G_1 \otimes G_2$. Both aggregated graphs $G_1$ and $G_2$ may have a significant community structure, thus the community detection on both aggregations will provide interesting, distinct insights on the underlying fine-grained network. 

A real-life analogy would involve, for instance, aggregating a social network according either to geographical location (e.g., counties), or to age class: both may exhibit relevant community structures explaining on the one hand which counties interact together, and the other hand which age classes interact together. Both community structures can be lifted back to the social network. If all age groups are equally present in each location, then those two partitions of the social network, although both interesting in their own rights, are `orthogonal' to each other as in the examples above. Thus at least one of them will differ sharply from communities found directly on the social network.


In summary, different aggregations of the original network may induce community structures on the original network that are completely disaligned with one another, without necessarily being `wrong',  and that are either similar or dissimilar to the community structure computed directly on the original graph. 

\section*{The aggregability index}

Between the two extremes situations where the aggregating partition is completely aligned with the community structure in $G$, or completely orthogonal to it, one finds intermediate situations where node aggregation is expected to perturb more or less the community detection. 

We propose a metric capable of capturing to what extent node aggregation will preserve community detection by introducing the \emph{aggregability index}, $\eta$, as the fraction of information required to identify the community of a randomly chosen node, that is provided by the knowledge of its aggregation class:
\begin{equation}
\eta =  \frac{I(\mathcal{C} ; \mathcal{P})}{H(\mathcal{C})}.
\label{eq1}
\end{equation}
Here $H(\mathcal{C})$ is the Shannon entropy of the community partition, defined in the following way. As a thought experiment, pick a node uniformly at random in $G$. The community of the node is a random variable with Shannon entropy $H(\mathcal{C}) \triangleq - \sum_{C\in \mathcal{C}} P(C) \log P(C)$, with probability  $P(C)$ of a community $C$ being proportional to its number of nodes. Similarly, $I(\mathcal{C} ; \mathcal{P})$ is the Shannon mutual information between the community in the partition $\mathcal{C}$ and the random aggregation class in $\mathcal{P}$ of a randomly picked node of $G$. 

Our newly-defined aggregability index, $\eta$, ranges from $0$  to $1$. In the $\eta=0$ limit, the aggregation classes are independent from the communities in $\mathcal{C}$, which implies  that each node is aggregated with nodes from other communities. In particular, the community structure $\mathcal{C'}$ that we may compute in the aggregated network, once lifted back to the initial graph $G$, form communities which are unions of aggregation classes, thus also independent from the communities in $\mathcal{C}$. In short, using the notations above, we can write $I(\mathcal{C} ; f^{-1}(\mathcal{C'}))=0$.   

In the $\eta=1$ limit, the aggregation classes are subset of the communities, thus any edge-counting criterion will preserve the community structure. In short, we write $\mathcal{C} = f^{-1}(\mathcal{C'})$.
 
Between these extreme situations, the lifted communities $f^{-1}(\mathcal{C'})$ are neither independent nor fully aligned with $\mathcal{C'}$. In this case, we observe, due to the fact that $f^{-1}(\mathcal{C'})$ is a coarser partition than the aggregation partition $f^{-1}(\mathcal{C'})$, that $I(\mathcal{C} ; f^{-1}(\mathcal{C'})) \leq I(\mathcal{C} ; \mathcal{P})$ (in application of the so-called data-processing inequality in information theory). In summary, we have in all cases:
 
\begin{equation}\label{eq:eta-vs-I}
 	 \eta \geq \frac{I(\mathcal{C} ; f^{-1}(\mathcal{C'})) }{H(\mathcal{C})},
\end{equation}
which confirms $\eta$ as a `best-case' estimate of the closeness between the community structure on the original graph $G$ and its aggregation $G'$. 
 
We may relax (\ref{eq:eta-vs-I}) to make it more symmetric in $\mathcal{C}$ and $\mathcal{C'}$, by increasing the denominator:
 
\begin{equation}\label{eq:eta-vs-I-2}
 \eta \geq \frac{I(\mathcal{C} ; f^{-1}(\mathcal{C'})) }{H(\mathcal{C})+H(f^{-1}(\mathcal{C'}))},
 \end{equation}
 which can be written equivalently as  
\begin{equation}\label{eq:eta-vs-NMI}
\eta \geq \frac{1}{2} \text{NMI}(\mathcal{C},f^{-1}(\mathcal{C'})),
\end{equation} 
 where NMI denotes a popular way to measure the similarity between two partitions, explained in the Methods (see Eq. \ref{eq:NMI}).
 Note that if the aggregating partition is very coarse, with a few large aggregation classes, then we expect that $H(f^{-1}(\mathcal{C'})) \ll H(\mathcal{C})$, and 
 Eq. \ref{eq:eta-vs-I-2} is not more conservative than Eq. \ref{eq:eta-vs-I}. If, on the other hand, we only have a few nodes in each aggregation class and $H(\mathcal{C})$ is relatively large then we may heuristically expect $H(\mathcal{C}) \approx H(\mathcal{C'})$, and it is more relevant to write
 \begin{equation}\label{eq:eta-vs-NMI-2}
 \eta \gtrsim  \text{NMI}(\mathcal{C},f^{-1}(\mathcal{C'})).
\end{equation}

There is no reason that these inequalities should be always tight. Assume for instance that exactly one aggregation class overlaps over two communities $C_1$ and $C_2$ (so that $\eta <1$, if only by a little). Then in the aggregated network, the node resulting from this aggregation class will create edges whose weight typically depends on the \emph{density} of the two communities $C_1$ and $C_2$. Thus if $C_1$ and $C_2$ are sparse enough, the links so created in the aggregated network may be negligible so that the optimal community structure will not be modified, and the ideal situation 
$\mathcal{C} = f^{-1}(\mathcal{C'})$ that holds for $\eta=1$ and edge-counting criteria still holds. If on the other hand, the aggregation class cuts into dense communities, this will result in heavy weights in the aggregated that might disrupt significantly the overall community structure. We expect therefore that a network that is heterogeneous in terms of density of links may be potentially more fragile to aggregation, in terms of community structure. In section SA.2. of the Supplementary Information, we investigate the behaviour of the aggregability index $\eta$ in synthetic graphs with planted communities of heterogeneous densities.

In the next sections we show empirically how the aggregability index $\eta$ correlates with the NMI distance between the optimal partitions found for the original and aggregated networks on two datasets.  Albeit embedded in the same geographical area ---Belgium--- these two case studies will reveal different behaviours with respect to aggregation. In both cases, we 
know a network $G$, aggregate it according to administrative units or regular squares, compute the aggregability index and observe the distorsion of the communities found to be optimal in the new (aggregated) networks.

\section*{Methods}

We now describe the datasets, the definition of community and the way to compare partitions in an empirical approach. Both datasets are localised on parts of Belgium. See section SA.1. of the Supplementary Information for a visualisation and description of the territory.

\subsection*{Twitter networks}

Our first dataset is composed of 291,552 tweets geolocalised on the Belgian territory  between 18,327 Twitter users, obtained as described in Supplementary Information SA.2.  From this dataset we build a network $N_0$ as follows. The nodes are the users, and the weighted edges count the number of reply-to tweets between the two users (without taking the directionality into account, in order to keep the graph undirected). Each node is associated to a position, obtained as the barycentre of positions of the user recorded in each sent tweet.  In this way we see $N_0$ as a network linking positions together. By the means of how the dataset was collected, those positions are spread over the Belgian territory. 

A list of aggregated networks was created from  $N_0$. The territory of Belgium is divided into 589  municipalities, and used to be divided into 2,675 smaller municipalities until a merge took place in 1979. We first build two aggregated versions, where nodes represent former ($N_{fm}$) and current ($N_{m}$) municipalities, respectively, by merging all nodes of $N_0$ positioned in the same (former or current) municipality. Edges are merged accordingly, receiving a weight that aggregates the weights of all corresponding edges of $N_0$.
 
We also applied a regular grid of 125 m square cells onto the Belgian territory, and merged into a single node all nodes of $N_0$ positioned in the same cell, creating the aggregating network $N_{125}$. Increasingly coarser square grids of cell size 250 m to 32 km, were used in the same way to create the aggregated networks $N_{250}$ to $N_{32k}$ respectively. The number of nodes and edges are described in Table S1 of the Supplementary Information (SA.3.).  

\subsection*{Phone networks}

Our second dataset counts the numbers of phone calls between towers in the territory of Brabant, a former administrative unit (province) of 111 municipalities including and surrounding 
Brussels, the capital of Belgium. The derived undirected network, called $M_0$, is composed of 1,168 nodes (towers). A weighted edge between two towers counts the number of communications between the towers in either direction, for a total of 13M communications over the network. As each tower is associated with a precise position, one may again consider $M_0$ as a network between places. We may aggregate those places into municipalities, thus forming the network $M_m$, or into cells of regular size 125 m to 32 km, creating the networks $M_{125}$ to $M_{32k}$, as for the Twitter dataset. See Table S2 of the Supplementary Information, section SA.3, for the number of nodes and edges of each network.

\subsection*{Linearised stability maximisation}

Communities are intuitively meant here as sets of strongly interconnected nodes with comparatively few connections between the communities. Among the many formalisations of this concept, one of the most popular is modularity \cite{newman2004detecting}, quantifying the goodness of a given partition $\mathcal{C}$ of nodes as
\begin{equation}
Q_\mathcal{C}=\frac{1}{2m} \sum_{C \in \mathcal{C}} \sum_{i,j \in C} (A_{ij} - \frac{k_i k_j}{2m}),
\end{equation}
where $m$ is the sum of all weights of the networks' edges, and $k_i$ represents the (weighted) degree of node $i${.} $A_{ij}$ is the weighted adjacency matrix of the network, and 
$C (\in \mathcal{C}$) represents a community of the partition. 

We use a generalisation, called linearised partition stability \cite{delvenne2013stability}, or equivalently Potts model ~\cite{reichardt2007partitioning}, which introduces a resolution parameter $\rho$ varying from $0$ to $\infty$ as follows:


\begin{equation}
r_{lin}(\rho,\mathcal{C}) = (1-\rho) +   \rho \frac{1}{2m} \sum_{C \in \mathcal{C}} \sum_{i,j \in C} ( A_{ij} - \frac{1}{\rho}  \frac{k_i k_j}{2m}), 
\end{equation}

At $\rho=0$, single nodes are optimal as communities, while  partitions with larger communities emerge for increasing values of $\rho$, until a single community is 
optimal at  $\rho \to  \infty$. For $\rho=1$, the linearised stability is the modularity, $r_{lin}(1,\mathcal{C})=Q_\mathcal{C}$. 
The resolution parameter $\rho$ is hereafter
called timescale, because linearised stability is formally derived in \cite{delvenne2013stability} as capturing the ability of incumbent communities to retain the flow of a 
diffusion of random walkers across the network for a timescale of the order of $\rho$. The original Potts model~\cite{reichardt2007partitioning} uses the parameter $\gamma = 1/\rho$.
As most community detection criteria, linearised stability is NP-hard to optimise except 
for extreme values of $\rho$, and we use the Louvain method \cite{blondel2008fuc,lambiotte2014random} as a heuristic.

Whenever appropriate, we will use the linearised stability method to detect communities, because it is an edge-counting criterion, because it includes an extremely popular criterion (modularity, for $\rho=1$) as a special case, and because it allows adapting the timescale parameter $\rho$ in order to create partitionings on different networks with the same or similar number of communities. There are certainly many methods of merits sharing the same properties. Our goal in the Results section is not to find the most sociologically relevant Twitter or phone call communities in Belgium, but illustrate how partitions found with an edge-counting criterion are modified in presence of aggregation.  Therefore, the various arguments in favor or against the practical significance of the communities delivered by one or another method are not relevant here.     

\subsection*{Normalised mutual information for comparing partitions}

We compute the normalised mutual information \cite{ana2003robust}, between the two partitions $\mathcal{C}$ and $\mathcal{D}$ of the same set of nodes, to evaluate how similar they are, as 
\begin{equation} \label{eq:NMI}
\mathrm{NMI}(\mathcal{C},\mathcal{D})= \frac{I(\mathcal{C};\mathcal{D})}{( H(\mathcal{C})+H(\mathcal{D}) )/2},
\end{equation}
where $I(\mathcal{C};\mathcal{D})$ denotes the mutual information between the two partitions, i.e. between the set in $\mathcal{C}$ and the set in $\mathcal{D}$ containing a randomly picked node of the graph. Note that in this article, the sets of nodes belonging to a partition are either called `communities' (if found by community detection algorithm) or `aggregation classes' (if defining a way to aggregate the network).

 Similarly, $H(\mathcal{{C}})$ or $H(\mathcal{{D}})$ denotes the Shannon entropy of each partition, i.e., the Shannon entropy of the set of a randomly picked node 
of the graph. The NMI takes values between $0$, for independent (thus maximally dissimilar) partitions, and $1$, for identical partitions.

In our case, we also want to be able to compare community partitions at different levels of aggregation, let us say for example the optimal partition $\mathcal{C}$  and $\mathcal{D}$ of networks
$N_0$  and $N_{125}$, respectively. In this case, we lift the communities of $N_{125}$ into communities of $N_0$, replacing each node of $N_{125}$ by its aggregation classes in 
$N_0$.  We call $\mathcal{D}'$ this partition of the nodes of $N_0$.
 We now compare the two partitions $\mathcal{C}$ and $\mathcal{D}'$ with the quantity $\mathrm{NMI}(\mathcal{C},\mathcal{D}')$, which we will also sometimes denote $\mathrm{NMI}(\mathcal{C},\mathcal{D})$ by abuse of notations.

\section*{Results}

In the following, we illustrate on the two real-life datasets the concepts explained above on toy networks. Specifically, we show how the aggregation process over the Twitter and phone call networks strongly affects the community partition in the former case, and mildly so in the latter. 
We also show how the magnitude of this distorsion, as the aggregation grid becomes coarser and coarser, correlates with the proposed \emph{aggregability index}. 

\subsection*{Twitter networks}
\begin{figure}[h!]
\begin{center}
  \subfloat[Communities detected in the network of municipalities $N_m$]{\label{fig:comuni}\includegraphics[width=0.47\textwidth]{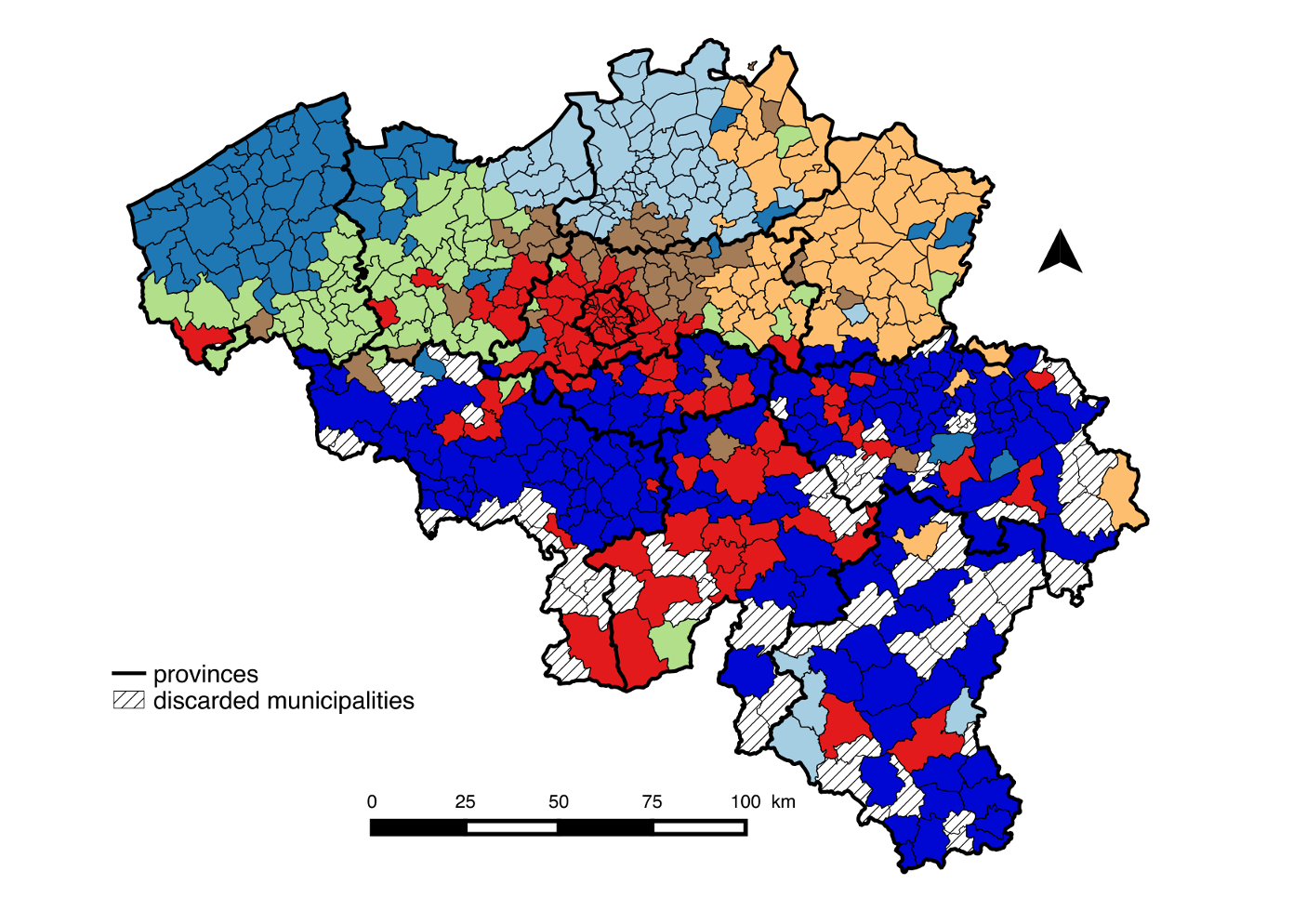}}
  \hspace{5pt}
  \subfloat[Community 1, network $N_0$]{\label{fig:comu1}\includegraphics[width=0.47\textwidth]{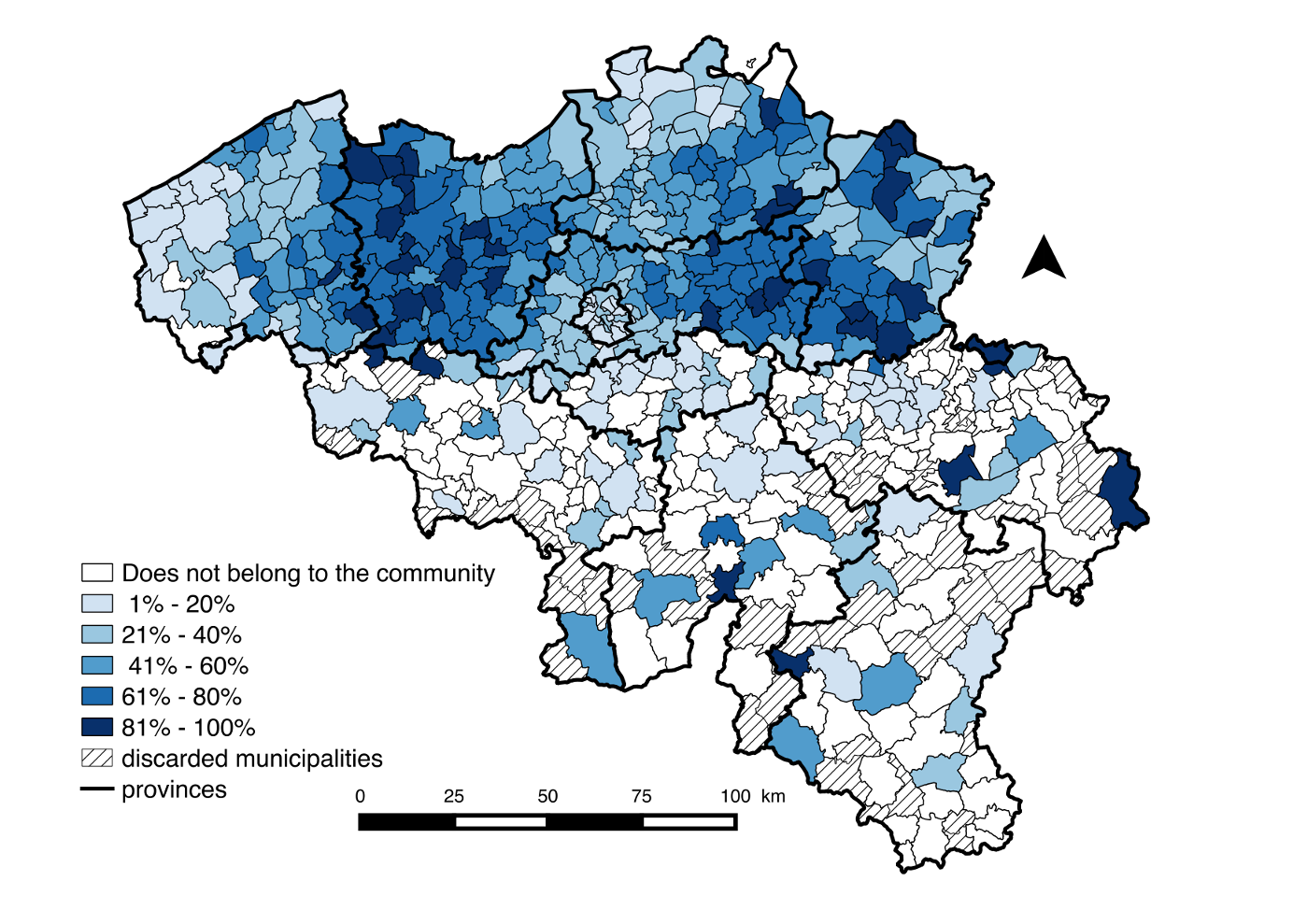}}
  \hspace{5pt}
  \subfloat[Community 2, network $N_0$]{\label{fig:comu2}\includegraphics[width=0.47\textwidth]{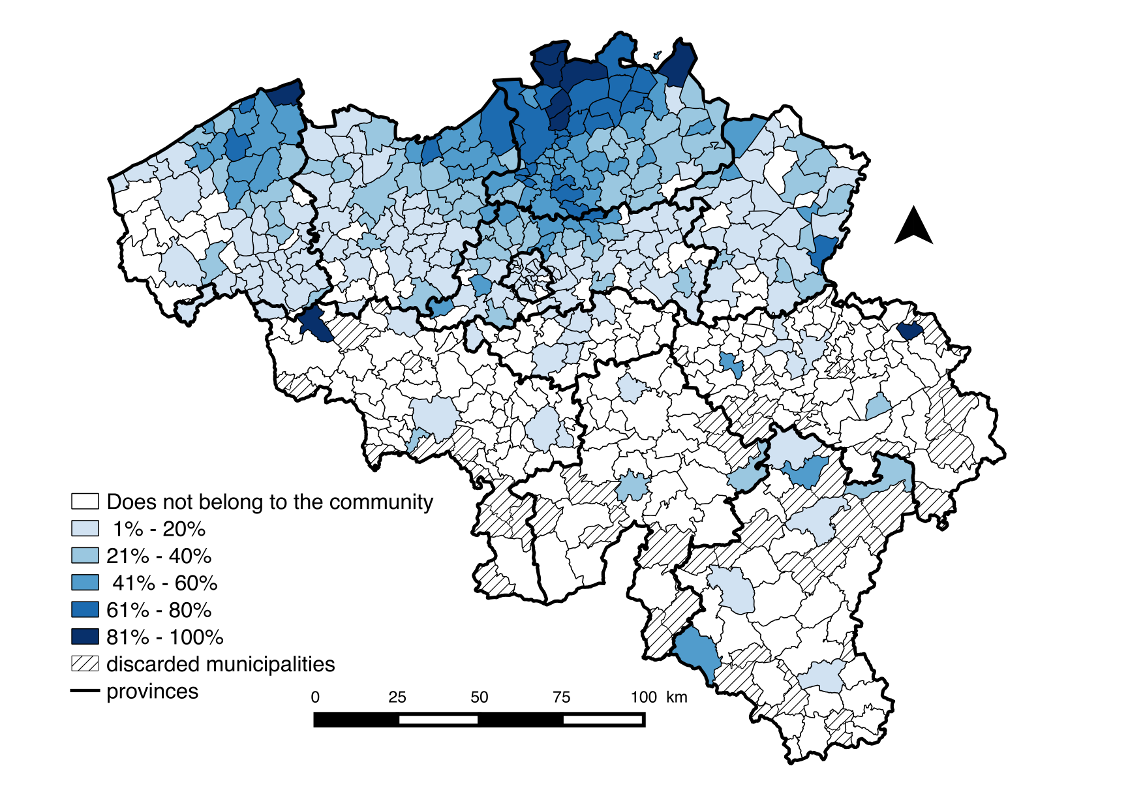}}
   \hspace{5pt}
  \subfloat[Community 3, network $N_0$]{\label{fig:comu3}\includegraphics[width=0.47\textwidth]{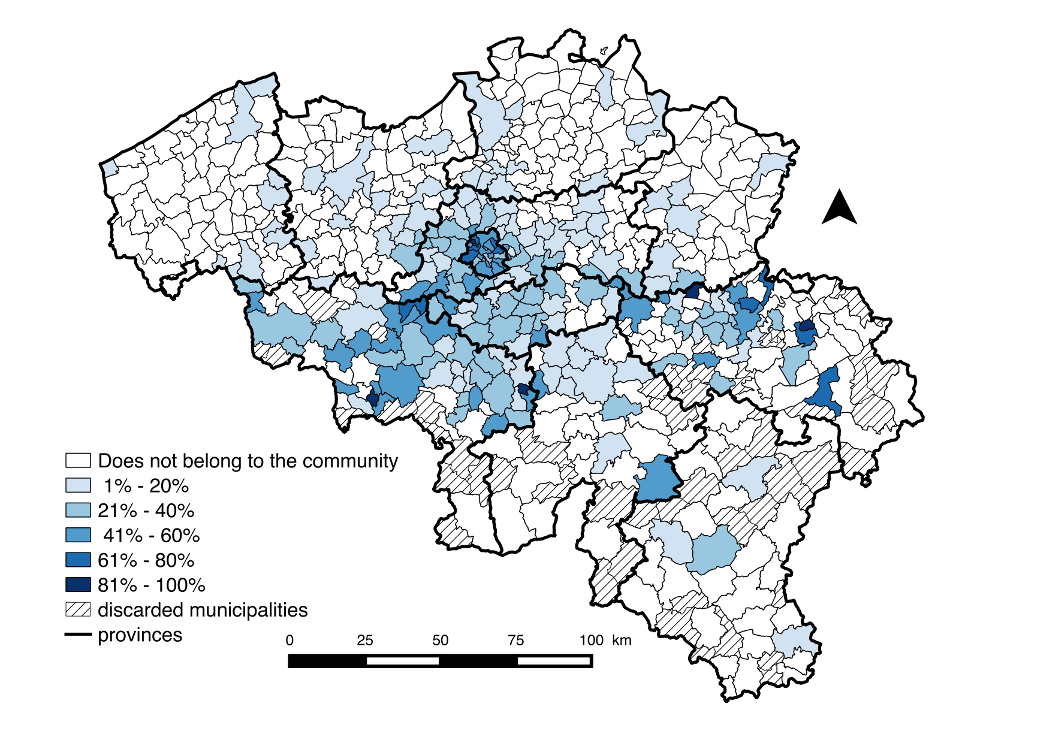}}
   \hspace{5pt}
  \subfloat[Community 4, network $N_0$]{\label{fig:comu4}\includegraphics[width=0.47\textwidth]{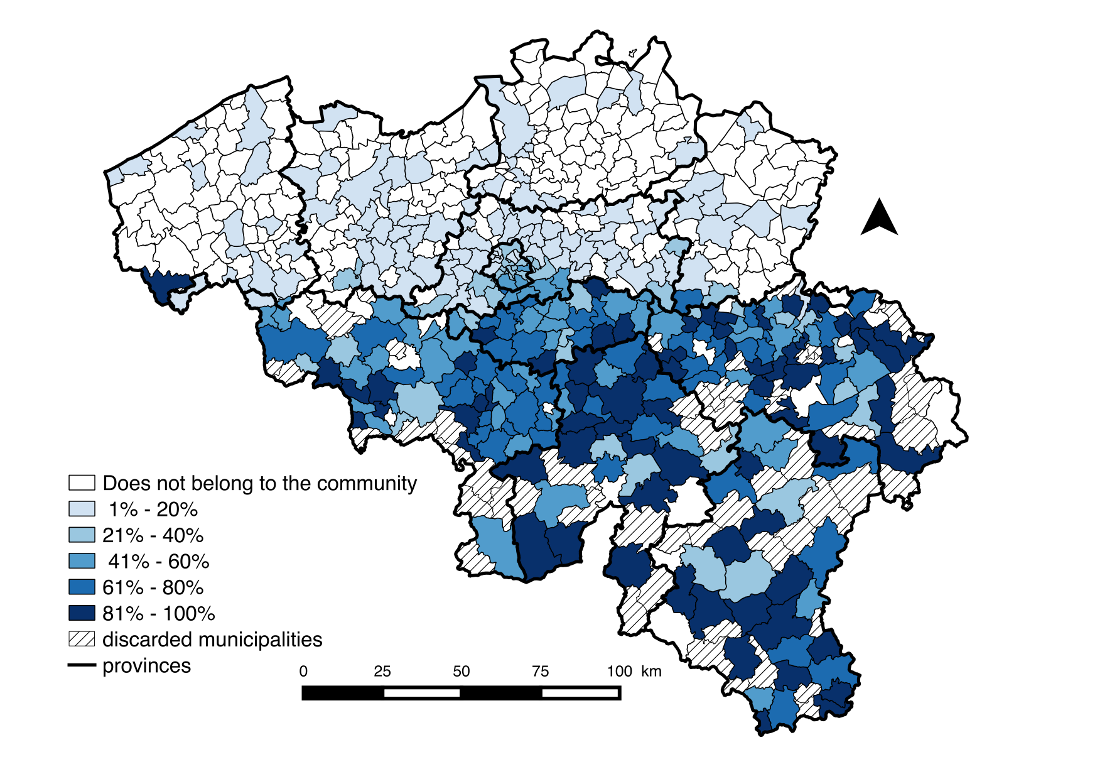}}
   \hspace{5pt}
  \subfloat[Community 5, network $N_0$]{\label{fig:comu5}\includegraphics[width=0.47\textwidth]{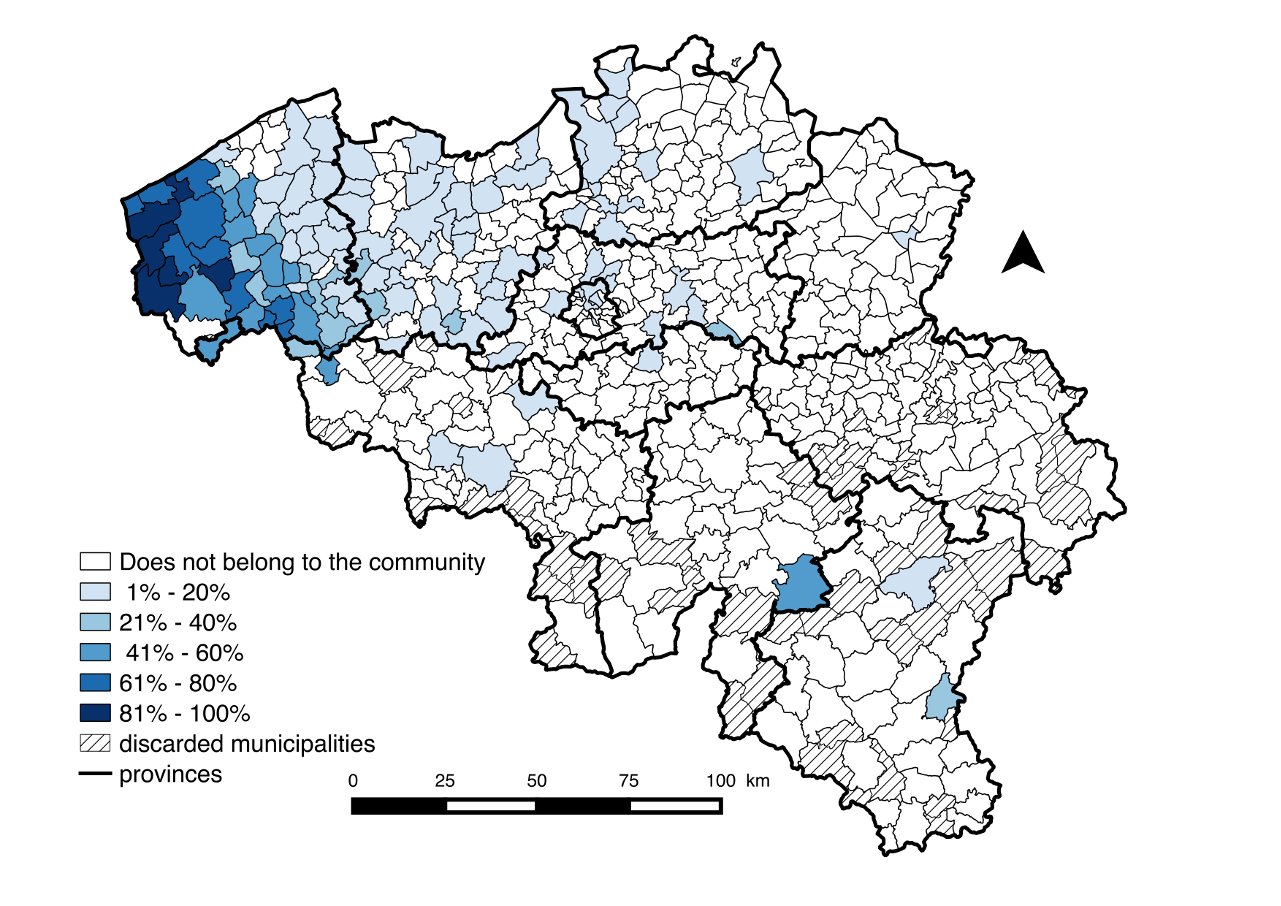}}
  \end{center}
  \caption{Spatial footprint of communities detected in Twitter networks $N_m$ (municipalities, timescale $\rho=1$) and $N_0$ (individual users, timescale $\rho=10$).}
  \label{fig:comgen}
\end{figure}

Figure \ref{fig:comgen}-a shows the communities extracted from the network $N_m$ of municipalities, using a timescale $\rho=1$. Each figure from \ref{fig:comgen}-b to 
\ref{fig:comgen}-f shows the spatial footprint of one community of individual Twitter users. We have used a timescale $\rho=10$, in order to illustrate the case with a number of communities (namely 5) comparable to the 7 communities of the  $N_m$ network. The colour intensity in each municipality represents the proportion of users positioned in this municipality  who belong to the community being represented. 

Some communities of $N_0$ (for example those represented on Figures \ref{fig:comgen}-b and \ref{fig:comgen}-c) show a remarkable geographical dispersion, and in particular do not seem to match any community of $N_m$ (only community $4$ in Fig.\ref{fig:comgen}-e seems to match a community in the network of municipalities in Fig.\ref{fig:comgen}-a, namely the dark blue one). 

In order to analyse quantitatively the effect of aggregating data, we systematically test different levels of spatial aggregation, all at the same timescale parameter $\rho=1$. In other words, for the next analysis, we look at the maximum modularity communities, as approximated by the Louvain method. 

\begin{figure}[h!]
\begin{center}
\subfloat[Network $N_{1km}$, cells of side 1km]{\label{fig:grid1}\includegraphics[width=0.47\textwidth]{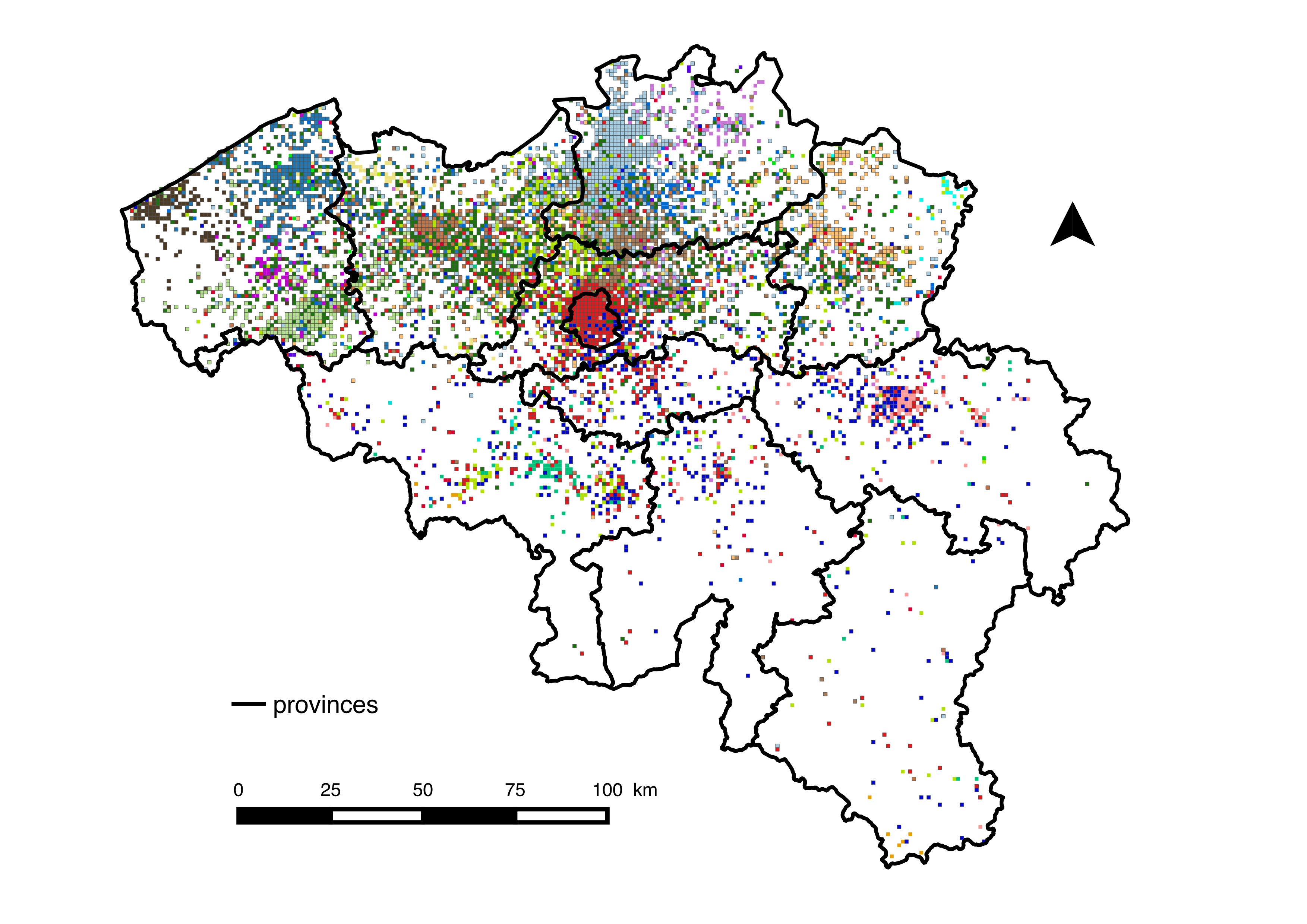}}
\hspace{5pt}
\subfloat[Network $N_{2km}$, cells of side 2km]{\label{fig:grid25}\includegraphics[width=0.47\textwidth]{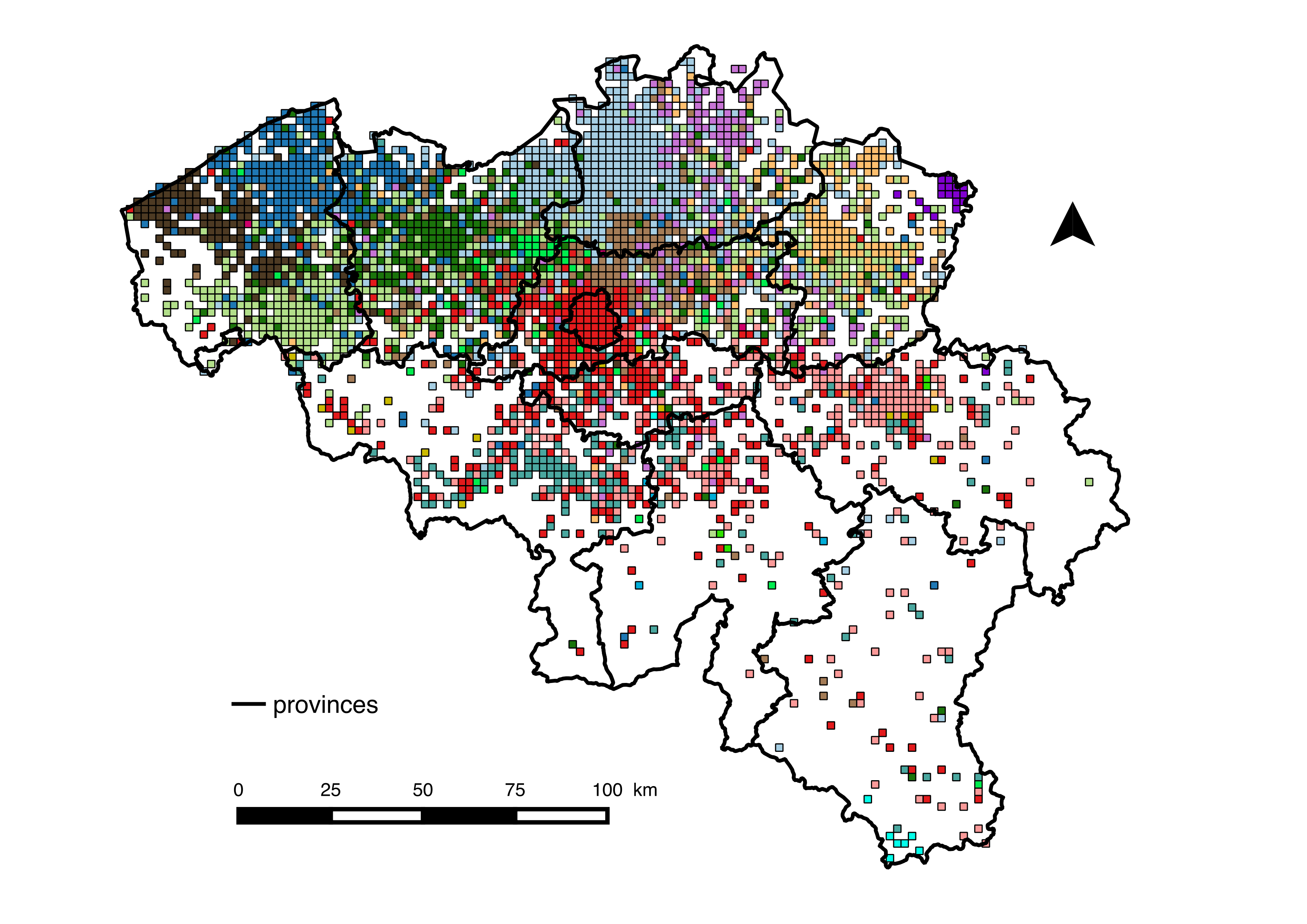}}
\hspace{5pt}
\subfloat[Network $N_{fm}$]{\label{fig:formermuni}\includegraphics[width=0.47\textwidth]{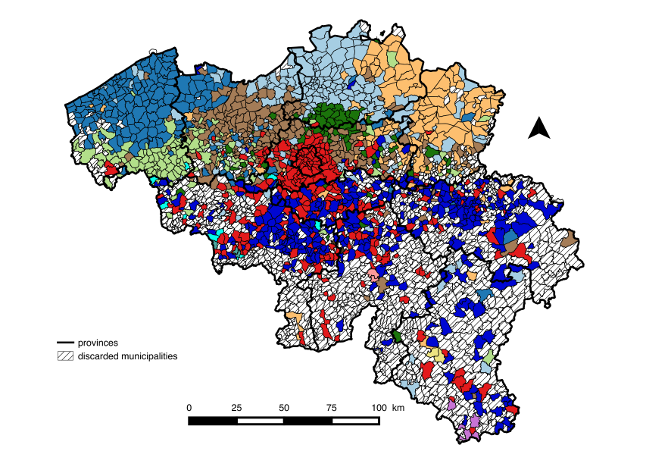}}
\hspace{5pt}
\subfloat[Network $N_{4km}$, cells of side 4km]{\label{fig:grid5}\includegraphics[width=0.47\textwidth]{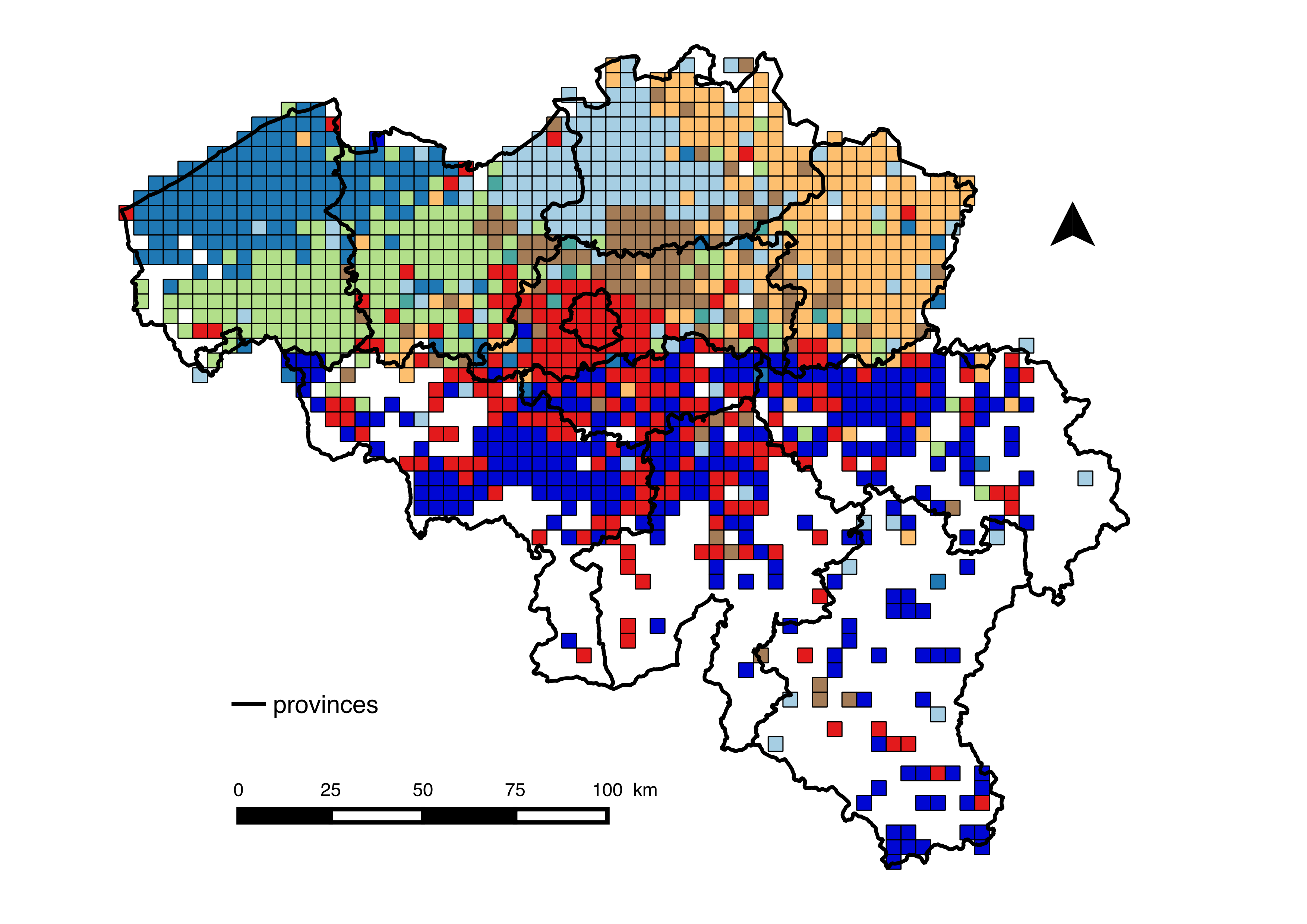}}
\hspace{5pt}
\subfloat[Network $N_{m}$]{\label{fig:grid5}\includegraphics[width=0.47\textwidth]{repto_com_0426.png}}
\hspace{5pt}
\subfloat[Network $N_{8km}$, cells of side 8km]{\label{fig:grid10}\includegraphics[width=0.47\textwidth]{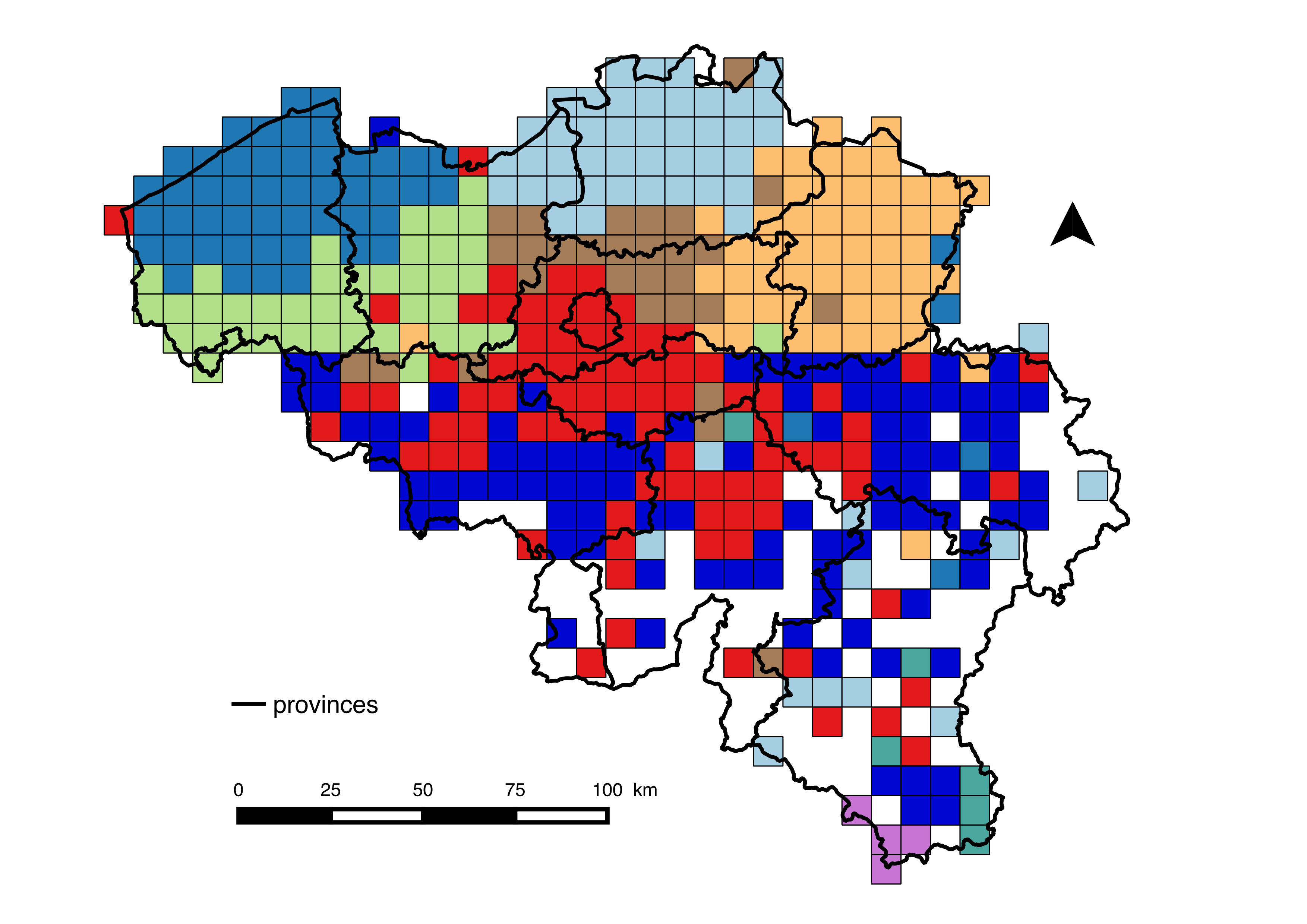}}
\end{center}
\caption{Communities detected in the Twitter network aggregated at the level of former municipalities, $N_{fm}$, (c), at the level of current municipalities in Belgium, $N_{m}$, (e), 
and aggregated into grids of square cells of different sizes (a-b, d, f) (timescale parameter set to 1).}
\label{fig:formermunigrids}
\end{figure}

Figure \ref{fig:formermunigrids} shows communities at different levels of aggregation: municipalities, former (smaller) municipalities and square cells of size 1km, 2km, 4km, 8km. As the aggregation classes become larger and larger they step over several communities forcing their re-arrangement and giving rise to different partitionings. 

We can see that as the nodes are increasingly aggregated, some communities gathering distant places, such as the light green community  in Fig. \ref{fig:formermunigrids}-a) to \ref{fig:formermunigrids}-c), are re-arranged into geographically localised communities (light green in Fig. \ref{fig:formermunigrids}-f). 

White areas represent the physical space where no event has been recorded. At the finest level ($N_0$), nodes are represented as a single point (the average position of a user), thus almost all space is white. As the aggregation scale increases, the white space is progressively removed, being merged with neighbouring space with non-zero activity. We observe that this 
effect is more visible in areas with low levels of activity, as the southern part of the country. 

The normalised mutual information (NMI) between the disaggregated network $N_0$ and several aggregated networks is depicted on Figure 
\ref{fig:nmi_evol}. Starting with the first level of aggregation (125 m), we observe that the NMI already drops rather steeply, even though
there is some fit (NMI $\approx 0.7$) between the communities displayed by aggregated units of 125 m and the non-aggregated ones. Values of NMI continue to  decrease with the size of the aggregation.

\subsection*{Mobile phone networks}
\begin{figure}[htp]
  \begin{center}
  \subfloat[Network of cell towers $M_t$]{\label{fig:mpcomuni}\includegraphics[width=0.47\textwidth]{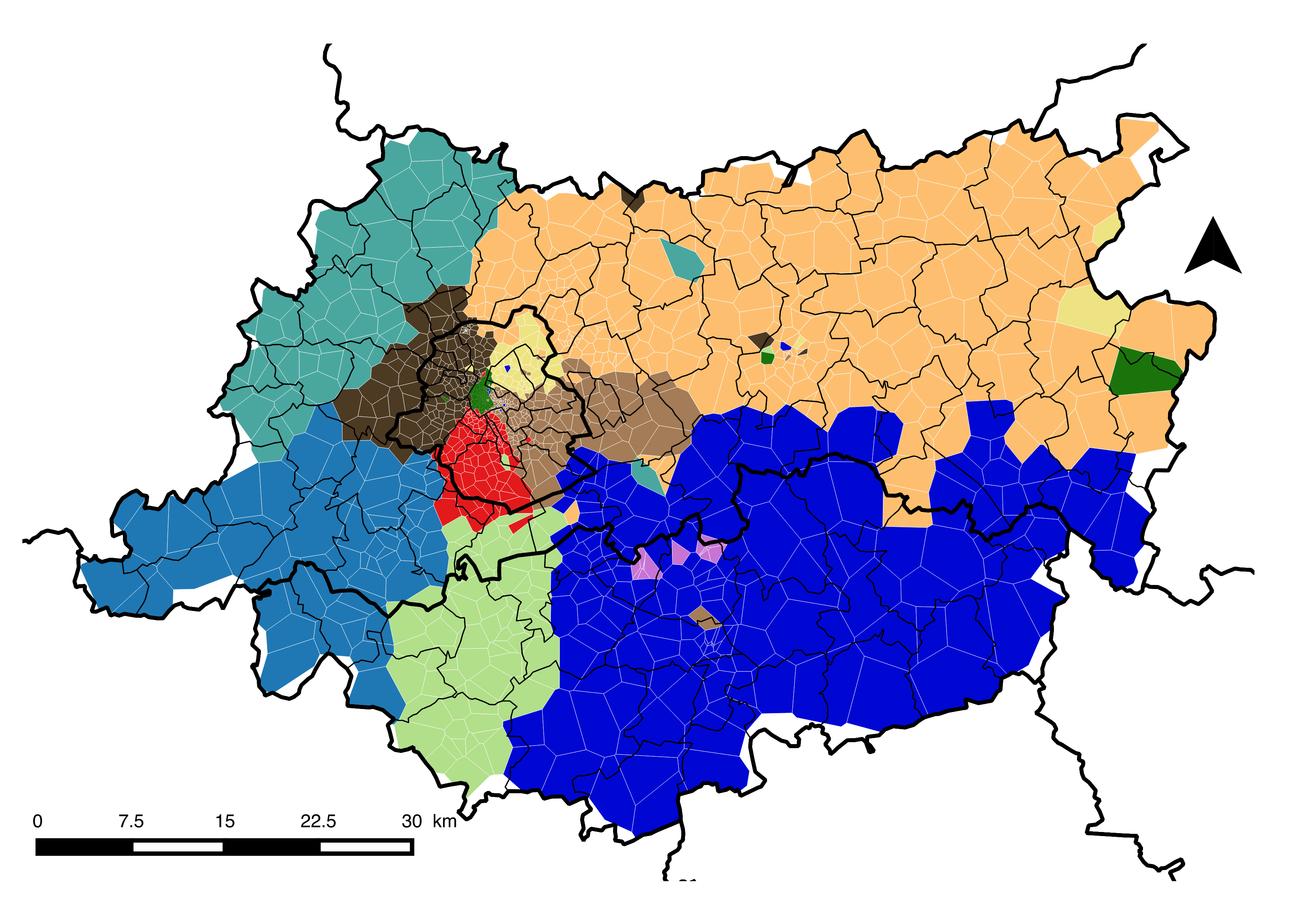}}
  \hspace{5pt}
  \subfloat[Network of communes $M_m$]{\label{fig:mpcomcell}\includegraphics[width=0.47\textwidth]{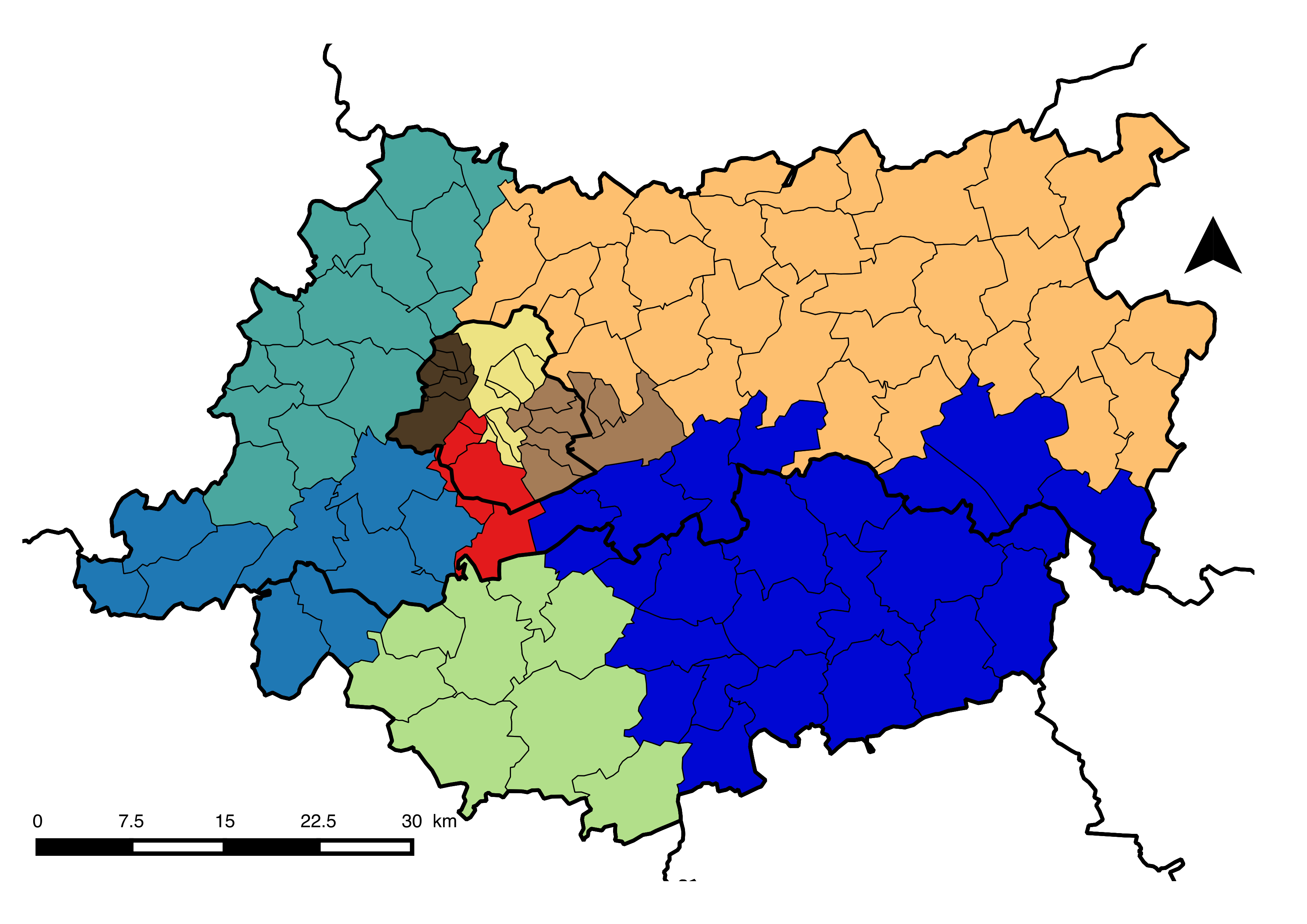}}
  \end{center}
  \caption{Communities detected in the mobile phone networks, at the disaggregated level of towers $M_0$ (note that although towers are characterised by a single point, for the visual depiction we represent them by the Voronoi polygone associated to it), and the aggregated level of municipalities $M_m$. The normalised mutual information, NMI, between partitions of network $M_0$ and $M_m$ is 0.64. Thus,  the similarity between the communities found on the two levels of aggregation is higher than the similarity observed in the Twitter network between the disaggregated network of users ($N_0$), and the aggregated versions (see Fig.~\ref{fig:nmi_evol}).   The timescale parameter $\rho$ is set to $0.75$, as suggested by another study on the same dataset \cite{thomas2017}. } 
  \label{fig:mpcomgen}
\end{figure}

Fig. \ref{fig:mpcomgen} shows the communities found at the disaggregated level of towers $M_0$ (note that although towers are characterised by a single point, for the visual depiction we represent them by the Voronoi polygone associated to it), and the aggregated level of municipalities $M_m$. The normalised mutual information, NMI, between community partitions found on networks $M_0$ and $M_m$ is 0.64. Thus,  the similarity between the communities found on the two levels of aggregation is higher than the similarity observed in the Twitter network between the disaggregated network of users ($N_0$), and the aggregated versions (see Fig.~\ref{fig:nmi_evol}). On Fig.~\ref{fig:nmi_evol} we also notice that the NMI between the communities found on $M_0$ and versions aggregated with larger and larger cells is consistently higher than in the case of the Twitter dataset.

\subsection*{Aggregability index and NMI}

\begin{figure}[h!]
\begin{center}
\includegraphics[width=0.7\textwidth]{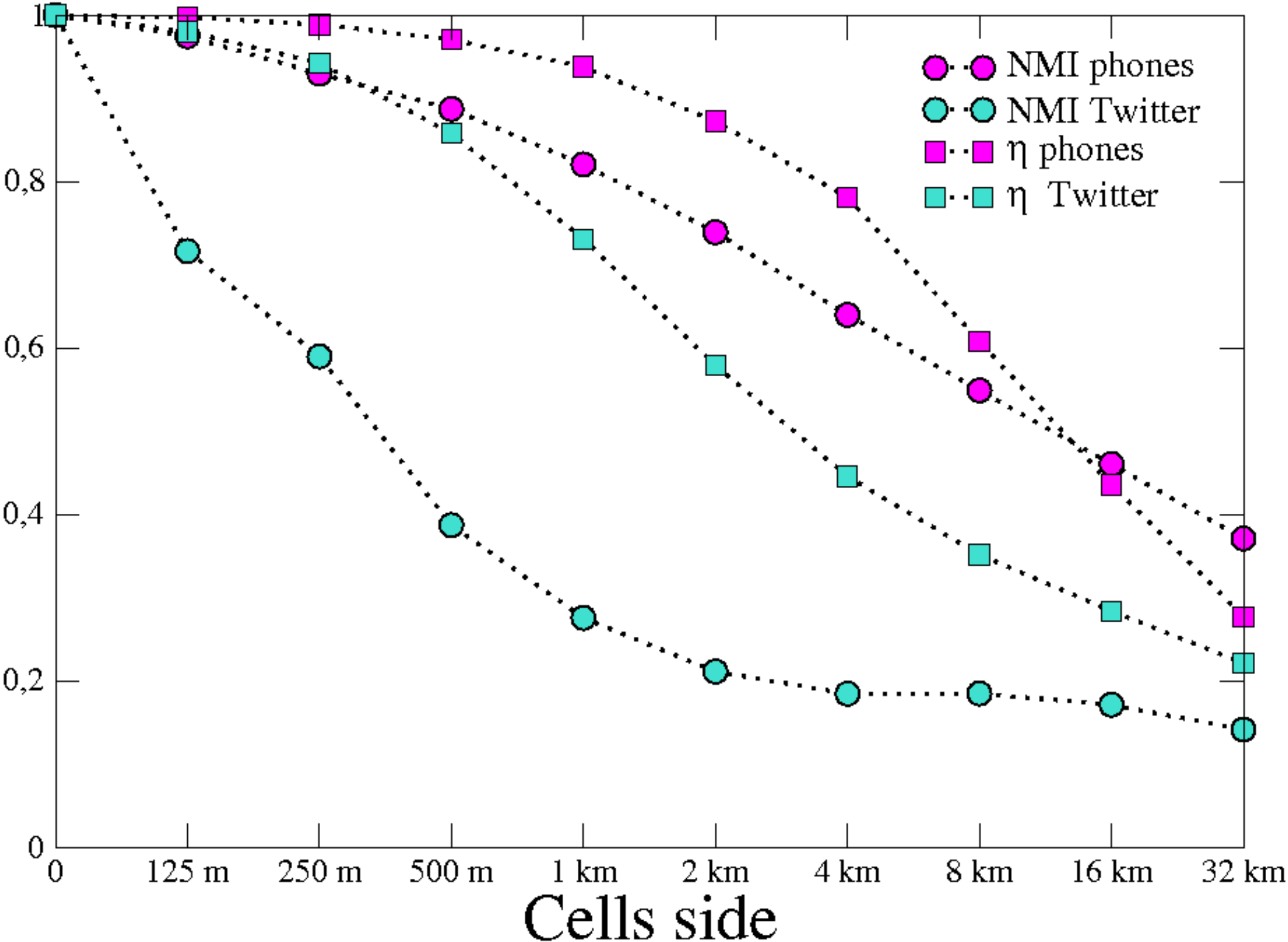}
\caption{In circles is shown the evolution of normalized mutual information, NMI, between communities found in the network prior aggregation, and communities found in aggregated 
networks at several square sizes. In squares, the evolution of the aggregability index, $\eta$, between communities and aggregability at the finest level compared with the 
same sizes as before. For Twitter data (in blue) the initial level corresponds to users centroids and time scale kept to $\rho=1$. For mobile phone data (in pink), the initial level corresponds to cell towers and the timescale was kept 
constant with a value of 0.75. }
\label{fig:nmi_evol}
\end{center}
\end{figure}

In Fig. \ref{fig:nmi_evol}, we compare for both datasets, the results of community detection on the original network ($N_0$ or $M_0$) with communities found on the networks of square cells of sides 125 m, 250 m, 500 m, 1 km, 2 km, 4 km, 8 km, 16 km and 32 km. We also plot the aggregability indices, comparing the community structure found on the original networks ($N_0$ or $M_0$) with the aggregating partitions into square cells of sides 125 m, 250 m, 500 m, 1 km, 2 km, 4 km, 8 km, 16 km and 32 km.  

The aggregability index, $\eta$, requires the knowledge of the  partition into communities and of the partition into aggregation classes at the finest level, but not of the communities that are deemed to be relevant to the aggregated graph. It measures to what extent every  aggregation class is a subset of a single community, which is a sufficient condition for the community structure to be left invariant by the (edge-counting) community detection method, as argued in this paper.

The shape of the $\eta$ and NMI curves in Fig. \ref{fig:nmi_evol} is in line with the following facts:
\begin{itemize}
	\item If $\eta=1$ then $\text{NMI}=1$ (because we use an edge-counting criterion for detecting communities),
	\item For a small aggregation scale we expect from Eq. \ref{eq:eta-vs-NMI-2} that $\eta \gtrsim \text{NMI}$, 
	\item  For further aggregation scales, we know from Eq. \ref{eq:eta-vs-NMI} that $\eta \geq \text{NMI}/2$. 
\end{itemize}

Low values of $\eta$ can be seen as a warning signal  that communities on the aggregated network (once lifted back to the original network) will necessarily be significantly different than the original communities. In Fig. \ref{fig:nmi_evol} we observe that the value of $\eta$ for mobile phone calls stays remarkably steady until the aggregation scale of 1 or 2 km, while the $\eta$ value for the Twitter dataset dips comparably faster --- and so does the NMI between the community partitions at different scales, as expected.

The fact that the NMI curve of the Twitter dataset drops significantly faster than the $\eta$ curve shows that  Eqs \ref{eq:eta-vs-NMI} and \ref{eq:eta-vs-NMI-2} need not be tight. In line with the arguments we discuss above (below Eq. \ref{eq:eta-vs-NMI-2}),  one reason for this discrepancy may be a strong heterogeneity of the data in terms of density, as Figs \ref{fig:comgen} and \ref{fig:formermunigrids} suggest. 

\section*{Discussion}

In this paper, we have studied the impact of  data aggregation on community detection in networks. We have shown on theoretical and empirical examples 
that  data aggregation can preserve the community structure, destroy it, or highlight another relevant community structure. We have identified a class of methods able to preserve the community structure whenever it is aligned with the aggregation classes. We have defined an aggregability index that measures how aligned the community structure is with the aggregation classes.

The article has been structured as a proof of concept. The examples have focused on the most standard notion of communities, as highly interconnected set of nodes. Communities were computed with one of the most popular quality functions, namely modularity and its multiscale extension.  We focused on aggregating  geographical coordinates into spatial units of increasing size, in line with the well-known Modified Areal Unit Problem in geography. 

Nevertheless, from the theoretical considerations, we see that the conclusions may be potentially relevant for different notions of partitioning (e.g. stochastic block modelling) with various aggregation criteria, according to any node metadata such as age, school, etc. 

Therefore, broadly speaking, we see our investigation as a warning to data scientists grappling with networks on several levels of aggregation. Our message being that the results of their analyses may depend starkly on the level and nature of the aggregation. 

We chose two datasets behaving differently with respect to aggregation, as an illustration for our proposed parameter, the aggregability index. The fact that these two datasets are geographic in nature is incidental in our study, whose scope includes in principle any kind of network and their aggregations. Nonetheless, this might indicate a potential privileged applicability to space-embedded networks. Example of networks embedded in space abound, and the interaction between their structure and the way they unfold in space has triggered some interesting developments, see for example \cite{mascolo_measuring_urban_social_diversity, rosvall_networks_and_cities}. As to explaining why the two datasets behave differently with respect to a same aggregation strategy, one can only formulate hypotheses, whose investigation is beyond the scope of this paper, and may involve the analysis of other datasets with other community detection methods on the same geographical area \cite{arnaudadamJGS}. While the mobile phone calls dataset is shaped by the condition of previous social interaction, this constraint is not present, or to a lesser extent, in the Twitter dataset. Further differences between the datasets include the heterogenous density of events in the Twitter network, 
and the different geographic area (Belgium or surroundings of Brussels). Even more importantly, the mobile phone network's nodes at the finest scale are towers, which already aggregate a large number of users. 

The present study is certainly not without caveats. For instance in many cases it may be that the full network is inaccessible to the measurement (such as in the human brain connectomes, only available under aggregated form), or too large for most community detection algorithms. In this case, a computation of the aggregation index $\eta$ may not be available. Also, in many cases the aggregated network is available with weights on the edges that do not represent the sum of all interactions between all nodes of the aggregation classes, but only a tresholded version of it, for instance.  Ways to cope with this may be a focus of further research.

\section*{List of abbreviations}
$m$ : Meters \\
MAUP : Modifiable Areal Unit Problem \\
NMI : Normalised mutual information \\
$N_0$ : Disaggregated network of Twitter users\\
$N_{fm}$ : Network of former municipalities \\
$N_m$ : Network of municipalities \\
$N_p$ : Networks of Twitter users aggregated into cells of size $p$ meters \\
$M_0$ : Disaggregated network of towers\\
$M_p$ : Networks of towers aggregated into cells of size $p$ meters \\

\section*{Competing interests}
The authors declare no competing interests.

\section*{Data availability statements}
The anonymised phone call datasets used in this paper cannot be made publicly available due to a privacy contract signed between the authors and the phone company in order to avoid privacy issues. The twitter dataset analysed during the current study are available from the corresponding author on reasonable request.

\section*{Author's contributions}
All authors contributed to designe the research, perform the research, write the manuscript. 
\section*{Acknowledgements}
 This work was supported by Innoviris (project Anticipate - Prospective Research 88 BRU-NET), Federation Wallonia-Brussels (Concerted Research Action ARC 14/19-060), and Flagship European Research Area Network (FLAG-ERA) Joint Transnational Call FuturICT 2.0.
 
 \footnotesize {

\end{document}